# Gender assignment in doctoral theses: revisiting Teseo with a method based on Cultural Consensus Theory

Nataly Matias-Rayme - nataly.matias.rayme@gmail.com
Universidad Politécnica de Madrid

Iuliana Botezan – ibotezan@ucm.es -
Universidad Complutense de Madrid

Mari Carmen Suárez-Figueroa - mcsuarez@fi.upm.es -
Universidad Politécnica de Madrid

Rodrigo Sánchez Jiménez – rodsanch@ucm.es -
Universidad Complutense de Madrid – **Corresponding author**

**Abstract**: This study critically evaluates gender assignment methods within academic contexts, employing a comparative analysis of diverse techniques, including a SVM classifier, gender-guesser, genderize.io, and a Cultural Consensus Theory based classifier. Emphasizing the significance of transparency, data sources, and methodological considerations, the research introduces nomquamgender, a cultural consensus-based method, and applies it to Teseo, a Spanish dissertation database. The results reveal a substantial reduction in the number of individuals with unknown gender compared to traditional methods relying on INE data. The nuanced differences in gender distribution underscore the importance of methodological choices in gender studies, urging for transparent, comprehensive, and freely accessible methods to enhance the accuracy and reliability of gender assignment in academic research. After reevaluating the problem of gender imbalances in the doctoral system we can conclude that it's still evident although the trend is clearly set for its reduction. Finaly, specific problems related to some disciplines, including STEM fields and seniority roles are found to be worth of attention in the near future.

# Introduction

The participation of women in science, the evolution of their presence in scientific production, the roles, the characteristics of their contributions to scientific work, and the features of gender gaps or inequalities associated with gender in science have been the subject of numerous studies over the years. Particularly, gender bias has received special attention from an early stage (Zuckerman and Cole, **1975**; Schiebinger, **1987**; Etzkowitz, et al., **1992**). In the last years, some ambitious studies have provided an overview of the prevalence of the issue of gender inequalities in science, its evolution, and its distribution across countries or disciplines (Larivière et al., **2013**; Holman, Stuart-Fox & Hauser, **2018**; Huang et al., **2020;** Andersen, **2023,** El-Ouahi & Larivière **2023**). Collaboration, research team structure and differences in authorship roles have been examined by several scholars (West et al. **2013**; Fell & König, **2016**; Macaluso **2016;** Abramo, D'Angelo & Di Costa, **2019**). Also, Differences in top performing tiers of authors have also been examined (Chan & Torgler, **2020**; Ioannidis et al., **2023**) and huge amounts of data have been scrutinized to generate a longitudinal perspective that allows for the consideration of full careers, which is a very promising approach (Huang et al. **2020**; Boekhout, Weijden & Waltman, **2021**).

Research on higher education, particularly focusing on the doctoral phase, has identified patterns that have relevant implications for the talent pool of the academia and can be related to the gender of supervisors and PhD candidates. Gaule & Piacentini (**2018**) found that homophily exists in supervisor/student associations and that it can be linked to increased entry rates into academia for new doctors, leading to the conclusion that the scarcity of same-gender advisors for female students might contribute to the under-representation of women in science and engineering faculty positions. Nicholas et al (**2017**) studied the role of gender in the perceptions of career progression among early career researchers, although they did not find that kind of association when interviewing a set of diverse background young scholars. This is not nevertheless the most common take on the issue, as (Borrego et al., **2010**; Reybold et al., **2012**) did detect a gender bias when studying postdoctoral scientists and their success rates when incorporating to academia. Also, Kim et al (**2022**) do find that women researchers have lower chances of joining the academia than men in every single field in their US study.

It is essential to identify individuals' gender to analyze and formulate strategies aimed at reducing the gender gap. Gender classification based on name labels has enabled scientific studies on gender that would otherwise be impractical, but its effectiveness is crucial for validating results. There is an ongoing debate regarding the validity and consequences of using the primary approach to the problem, gender classification based on given names. Gender categorization, often inferred from names and reduced to binary by algorithms, introduces hard-to-estimate biases in subsequent data processing steps (Mihaljević et al., **2019**). Gender identification based on names is specific to language and culture, as the accuracy of classification is higher when using a model trained with data from the target ethnicity (Malmasi & Dras, **2014**). This implies that the effectiveness of gender detection based on names can exhibit biases linked to countries of origin (Karimi et al., **2016**). Some names exhibit a 'weak genderization' depending on factors such as the geographical origin of the authors or their age, and even the existence of names that are used interchangeably for both women and men in the same cultural setting. Some of these issues are difficult to resolve, while others are inherently unsolvable. Although the frequency with which these types of names appear is very low, they require careful consideration.

In our exploration of gender assignment methods, our study pursues a dual objective. Firstly, we strive to move beyond manual procedures and scrutinize diverse alternatives, proposing a straightforward approach to assess them. To address this, we first conduct a comparative analysis of several methods employed to tackle such issues. These methods encompass a machine learning SVM technique, a paid API, and the Python packages gender-guesser and NQG. The aim is to ascertain which method exhibits superior effectiveness and to define the advantages or consequences of choosing one over the other.

Secondly, we intend to scrutinize how the adoption of more advanced assignment mechanisms influences the conclusions derived from studies utilizing less sophisticated techniques. To achieve this, we have revisited a database that has been subject to a recent study on the gender of academic staff. Teseo has served us as a specific case study, although we think that our conclusions could be useful to the study of gender in science as a whole. Finally, we have reexamined the issue of gender imbalance in the Spanish doctoral system using the data provided by the new method and contrasted our results with the findings of previous studies on this problem.

# Methods and data

To carry out the tests and select the most suitable method, four different procedures were chosen. Some of them incorporated well-developed and specific methods to properly preprocess the data, while others did not. Because of the latter, we utilized the Python package "nameparser", implemented by Gulbranson (**2023**) for straightforward analysis of human names into their individual components. The package relies on natural language processing (NLP) techniques and heuristic rules to analyze and break down full names into their individual components, such as last name, first name, and prefixes. It employs a pattern-based and regular expression approach to identify and extract specific parts of a name, considering common conventions and structures used in different cultures and regions.

## Cultural Consensus Theory Classifier- NQG

Van Buskirk, Clauset & Larremore (**2023**) propose a gender classification method based on names using an open-source cultural consensus approach. The method is based on the premise that certain names have a stronger cultural association with a particular gender. The method begins by collecting a list of gender-labeled names from different cultures and regions, integrating 36 different sources covering 150 countries and over a century. Then, a machine learning algorithm based on Bayes' theorem is used to calculate the relative frequency of each name associated with each gender on the list. Next, a cultural consensus approach is applied to determine the final gender assignment for a specific name. This approach estimates how and with what strength each name has gender, based on cultural associations reflected in a reference data corpus. The method has been used in other studies (LaBerge, **2024**) and even large-scale projects (Lin et al., **2023**), and has been validated by studies from other authors (Spoon et al., **2023**).

The aforementioned approach is implemented in a Python package called nomquamgender (NQG). NQG is a simple package that contains data and some functions to support gender classification based on names in scientific research. Computationally, this package provides access to name-gender association data that can be used to classify individuals into gender groups. However, this package does not make classifications itself but provides the probability that a name belongs to a female gender represented by p(gf). Names labeled with p(gf) smaller than 0.5 have been labeled male while names labeled above that threshold have been labeled as female.

| Name_label | parse_first_name | used | sources | counts | p(gf) | Gender |
|---|---|---|---|---|---|---|
| Sanchez Sanchez, Julio | Julio | julio | 27 | 136812 | 0.002 | Male |
| Perez Perez, Maria Victoria | Maria | maria | 34 | 1944793 | 0.953 | Female |
| Gomez Gomez, Elena | Elena | elena | 33 | 374211 | 0.992 | Female |
| Torres-Torres, Montserr | Montserr | montserr | 0 | 0 | | Unknown |
| Ruiperez-Perez, M. Mar | Mar | mar | 18 | 16305 | 0.722 | Female |

**Figure 1**: sample output from NQG and gender assignments.

**Figure 1** shows some examples, of which the first, "Julio Sánchez Sánchez" has a 0.2% probability of being female gender and has classified as male. The second, "Maria Victoria Pérez Pérez" has a 95% probability of being female gender and is classified as female. The fourth "Montserr Torres-Torres", on the other hand, shows an incomplete name, so it's not recognized by the method, and assigned to the Unknown category.

## Methods used for benchmarking

We used two methods that have been extensively used in the literature, albeit for other cultural settings, Gender-guesser and Genderize.io. Santamaría & Mihaljević (**2018**) provide a thorough evaluation of the methods using their test collection of international authors. The gender-guesser package, implemented by Saeta-Pérez (**2016**) relies on a statistical approach that uses historical name data to determine the probability that a given name belongs to a specific gender. However, it does not include routines that allow data preprocessing, so we used the Python library *nameparser* in order to detect given names. Genderize.io is a web service that can be reached via API, although it's not free. It has been implemented by the Danish company Demografix Aps, and is available online at https://genderize.io/. It predicts a person's gender using their given name and the statistical analysis of names and their association with genders. It uses machine learning algorithms and natural language processing techniques to make gender predictions.

A third method based on Support Vectors was used to complement the other two, as it was successfully used by other authors in the past (Malmasi & Dras, **2014**; Ghosh, **2022**). We used the Support Vector Classifier as implemented in the Scikit-learn (Pedregosa et al., **2011**), a widely used Python library for machine learning. To generate the necessary word embeddings, we used the method described by Leo (**2021**), and to train the classifier, the database published by Ghosh (**2022**) based on the names of Silicon Corporation's clients, with 84,899 records. These were divided into two parts, the training set (80% of the names) and the test set (the remaining 20%).

## Validation sets

We have utilized three distinct validation sets to assess the effectiveness of the four gender classification methods. The first set includes names of researchers from the University of Oviedo, a subset of the "Ranking of Researchers in Spain" compiled by Aguillo-Caño (**2022a**), for which the University of Oviedo (**2022**) added gender information. The second set encompasses the first 5000 researchers from the Ranking of Researchers in Spain, and the third comprises the combination of this set with an additional number of female researchers from a specialized female researcher ranking (Aguillo-Caño, **2022b**) to provide a different gender distribution. **Table 1** shows the distribution of gender across the three sets, as well as their size. The differences in size and distribution are meant to provide us with different environments that would depict a fairly similar distribution of genders in the Spanish academia (set 1), a more skewed distribution that favors males (set 2) and a distribution that favors females (set 3).

| Sets | Male | | Female | | Total |
|---|---|---|---|---|---|
| | Count | % | Count | % | Count |
| Set 1 | 707 | 60% | 474 | 40% | 1181 |
| Set 2 | 3896 | 78% | 1104 | 22% | 5000 |
| Set 3 | 3896 | 41% | 5677 | 59% | 9573 |

**Table 1**: Number of individuals and gender in each set

The first set includes gender labels that were provided inhouse by the staff at the University of Oviedo. We have double checked these labels, also using a manual procedure. The second set did not originally include gender labels, and the third set was partially labelled. Some of the labels could be obtained from the female researcher ranking (Aguillo-Caño, **2022b**). For those researchers not included in this ranking an individual search was conducted on Google Scholar.

Gender was determined manually based on photo and bio information of the profiles of researchers. The three sets are publicly available online[i].

## Validation metrics

**%Correct Female**: calculated as the percentage of female researchers that were correctly classified. The calculation is done by dividing the number of correctly classified female researchers by the total actual number of female researchers. According to **Table 2**, the formula for the %Correct Female metric is $TF/TF+FF$.

**%Correct Male**: calculated as the percentage of male researchers that were correctly classified. The calculation is done by dividing the number of correctly classified male researchers by the total actual number of male researchers. The formula for the %Correct Male metric is $TM/FM+TM$.

**Accuracy:** represents the proportion of correctly classified researchers, both male and female, relative to the total evaluated samples. The accuracy metric has been calculated as $TF+TM/TF+FF+FM+TM$.

**F1**: combines precision and recall into a single value that represents the balance between the model's precision and recall. Precision refers to the proportion of samples correctly classified as positive among all samples classified as positive, and recall refers to the proportion of true positive samples correctly classified as positive among all true positive samples. This would be formalized as $2×TF/((2×TF)+FM+FF)$.

|  | Prediction | |  |
| --- | --- | --- | --- |
| **Ground truth** | **Female** | **Male** | **Total** |
| **Female** | TF | FF | **TRF** |
| **Male** | FM | TM | **TRM** |
| **Total** | **PF** | **PM** | T |

**Table 2**: Confusion matrix model for the predictions of the different methods

# Results

As we have mentioned earlier, this work has various goals, being the first one to examine different options for gender assignment beyond manual procedures. After this we propose to evaluate the effect that more advanced assignment mechanisms could have on the conclusions drawn from studies using less sophisticated techniques. We have used Teseo as a use case, although undoubtedly the application of these approaches can be much broader, varied, and ambitious. We have also reevaluated the problem of gender imbalance in the Spanish doctoral system using new data. We divide the results section into three parts that deal with the three stated goals.

## Evaluation of methods

**Table 3** presents the percentage of labels with information on recognized first and last names for each method used in gender prediction. In general terms, 98% of the labels are ready to be assigned a gender, meaning that only 2% do not have a specific gender assigned. However, it is observed that the gender_guesser package, used for list 1, has 94% of labels ready to be assigned a gender. This is because some names are not registered in its database.

| Sets  | SVM  | Genderize.io | gender_guesser | NQG  |
|-------|------|--------------|----------------|------|
| Set 1 | 1,00 | 0,99         | 0,94           | 0,99 |
| Set 2 | 1,00 | 0,99         | 0,96           | 1,00 |
| Set 3 | 1,00 | 0,99         | 0,95           | 1,00 |

**Table 3**: Percentage of recognized labels for each method for gender prediction

Although the set of names that we are using is not particularly difficult, as Spanish names have a good coverage in many available training sets and reference lists, it's surprising that the SVM classifier apparently recognizes every label, reporting a 100% recall that has no match over the rest of the methods, although rates are very high in Genderize.io and NQG finds a correspondence for every name in the second and third sets.

|           | **Prediction - Set 1** |        |        |              | **Prediction - Set 1** |        |        |
|-----------|------------------------|--------|--------|--------------|------------------------|--------|--------|
| **SVM**   | **Female**             | **Male** | **Total** | *Genderize.io* | **Female**          | **Male** | **Total** |
| Female    | 157                    | 317    | 474    | Female       | 455                    | 9      | 464    |
| Male      | 220                    | 487    | 707    | Male         | 6                      | 694    | 700    |
| Total     | 377                    | 804    | 1.181  | Total        | 461                    | 703    | 1.164  |
| *g_guesser* | **Female**           | **Male** | **Total** | **NBG**    | **Female**             | **Male** | **Total** |
| Female    | 431                    | 8      | 439    | Female       | 464                    | 5      | 469    |
| Male      | 5                      | 664    | 669    | Male         | 6                      | 695    | 701    |
| Total     | 436                    | 672    | 1.108  | Total        | 470                    | 700    | 1.170  |
|           | **Prediction - Set 2** |        |        |              | **Prediction - Set 2** |        |        |
| **SVM**   | **Female**             | **Male** | **Total** | *Genderize.io* | **Female**          | **Male** | **Total** |
| **Female** | 372                   | 732    | 1.104  | Female       | 1.066                  | 12     | 1.078  |
| **Male**  | 1.280                  | 2.616  | 3.896  | Male         | 57                     | 3.814  | 3.871  |
| **Total** | 1.652                  | 3.348  | 5.000  | Total        | 1.123                  | 3.826  | 4.949  |
| *g_guesser* | **Female**           | **Male** | **Total** | **NBG**    | **Female**             | **Male** | **Total** |
| **Female** | 1.026                 | 13     | 1.039  | Female       | 1.099                  | 2      | 1.101  |
| Male      | 46                     | 3.697  | 3.743  | Male         | 63                     | 3.826  | 3.889  |
| Total     | 1.072                  | 3.710  | 4.782  | Total        | 1.162                  | 3.828  | 4.990  |
|           | **Prediction - Set 3** |        |        |              | **Prediction - Set 3** |        |        |
| **SVM**   | **Female**             | **Male** | **Total** | *Genderize.io* | **Female**          | **Male** | **Total** |
| Female    | 2.000                  | 3.677  | 5.677  | Female       | 5.521                  | 65     | 5.586  |
| Male      | 1.280                  | 2.616  | 3.896  | Male         | 57                     | 3.823  | 3.880  |
| Total     | 3.280                  | 6.293  | 9.573  | Total        | 5.578                  | 3.888  | 9.466  |
| *g_guesser* | **Female**           | **Male** | **Total** | **NBG**    | **Female**             | **Male** | **Total** |
| Female    | 5.271                  | 48     | 5.319  | Female       | 5.629                  | 30     | 5.659  |
| Male      | 46                     | 3.697  | 3.743  | Male         | 63                     | 3.826  | 3.889  |
| Total     | 5.317                  | 3.745  | 9.062  | Total        | 5.692                  | 3.856  | 9.548  |

**Table 4**: Confusion matrices for the three sets and four methods

**Table 4** presents the confusion matrix for each method for each of the available sets. This information can be used to verify our metrics and calculate new ones. It does not include an "unknown" class, but as the rates of unrecognized names were so low, the effect on the F1 metric would necessarily be very low.

|  | SVM | Genderize.io | *gender_guesser* | NQG |
|---|---|---|---|---|
| **Metrics** | Set 1 | | | |
| %Correct female | 0,33 | 0,98 | 0,98 | 0,99 |
| %Correct male | 0,69 | 0,99 | 0,99 | 0,99 |
| Accuracy | 0,55 | 0,99 | 0,99 | 0,99 |
| F1 | 0,37 | 0,98 | 0,99 | 0,99 |
| **Metrics** | Set 2 | | | |
| %Correct female | 0,34 | 0,99 | 0,99 | 1,00 |
| %Correct male | 0,67 | 0,99 | 0,99 | 0,98 |
| Accuracy | 0,60 | 0,99 | 0,99 | 0,99 |
| F1 | 0,27 | 0,97 | 0,97 | 0,97 |
| **Metrics** | Set 3 | | | |
| %Correct female | 0,35 | 0,99 | 0,99 | 0,99 |
| %Correct male | 0,67 | 0,99 | 0,99 | 0,98 |
| Accuracy | 0,48 | 0,99 | 0,99 | 0,99 |
| F1 | 0,45 | 0,99 | 0,99 | 0,99 |

**Table 5**: Performance metrics for the three sets and the four methods

As can be seen in Table 4, except for the Support Vector Machines-based classifier, the accuracy and correctness rates are very high. Even the F1 measure, which is a more severe metric, has very high scores. In our opinion, this difference is due to the different quality of the training data that the SVM classifier and the other methods have used. Both Genderize.io, gender_guesser and NQG use varied and large data sources, with broad coverage of various cultures and time periods.

A better data source that was correctly organized for use in the SVM classifier should improve performance, but gathering a wide array of appropriate sets from different cultures is not an easy task. This highlights the inherent advantage of using systems with access to a wide range of different sources, as methods for gender recognition/assignment are highly dependent on available information. On the other hand, among the remaining three methods, performance metrics are very similar. Of the three, gender_guesser is the one that recognizes a lower percentage of labels, but with a not very significant difference. It might be related to the fact that its data is clearly outdated, as it comes from a 2007 software development (Saeta-Pérez, **2016**). On the other hand, Genderize.io is a very effective system but offers unfavorable access conditions for even medium-sized projects. Under these conditions, nomquamgender offers the best balance of performance, breadth, and depth of data, and access conditions.

A relevant aspect of gender assignment that González-Salmón & Robinson-García (**2024**) highlight is that many services and software packages are not entirely transparent in revealing data sources on names and genders or are outright opaque. We believe that nomquamgender is precisely paradigmatic from the perspective of transparency and reproducibility, as it makes all datasets used during the name assignment process available to researchers (Van Buskirk, Clauset, & Larremore, **2023**). Considering these arguments, we decided to use NQG as a tool for assigning gender in the practical case presented below.

Updating gender data in Teseo dissertation database

Teseo is a dissertation database developed and maintained since 1976 by the Ministry of Universities of Spain. It contains information about doctoral candidates (successful), advisors, committee members, as well as data related to the thesis itself (including a classification of topics and an abstract), and information regarding the home institution of the doctoral program. Teseo is a data source regularly used in the Spanish context, serving as the basis for numerous studies on the production of doctoral theses across various fields of knowledge (Curiel-Marín & Fernández-Cano, **2015**; Ramos-Pardo & Sánchez-Antolín, **2017**), the relationship between thesis production and scientific output (Musi-Lechuga, Olivas-Ávila & Buela-Casal; Sánchez-Jiménez et al., **2017**) or the academics involved in the doctoral process (Olivas-Avila & Musi-Lechuga, **2010**; Repiso-Caballero, Torres-Salinas & Delgado-López-Cózar, **2011**). It has also been used to study gender in relation to dissertation production, the relationship between supervisor and the candidate, the inner workings of committees (Villarroya et al., **2008**; Prim-Espada, Diego-Sastre, Pérez-Fernández, **2010**; Ramos-Pardo & Sánchez-Antolín, **2017**; Hernández-González, Pano-Rodríguez, Reverter-Masia, **2019**; Maz-Machado et al., **2022**)

Sánchez-Jiménez et al. (**2023**) studied gender distribution by universities, scientific areas, and time periods to characterize gender inequalities and their evolution. They concluded that gender imbalance was still a relevant feature in their data and pointed to STEM disciplines and role differences as the most concerning issues about gender balance in the Spanish doctoral system. The original process of gender classification was based on a simple matching procedure that employed lists of 54,374 different names of people residing in Spain, and the frequency with which they were assigned to male and female genders. The lists are provided by the Spanish Statistical Office (INE, **2021**) and are freely accessible online. Records with incomplete or ambiguous information, as well as those that could not be analyzed due to errors in label structure elevated the percentage of academics with unknown gender to 18.2%. There's a possibility that this data fraction may not be uniformly distributed by gender but rather influenced by inherent biases in the data or the social and cultural conditions of its origin. This raises concerns about the reliability of such studies under conditions of high uncertainty (Boyack, **2023**). In a study like this, the reliability of the method used should be crucial, so it seems reasonable to check whether using a more advanced method than the one proposed by the authors can bring changes to the calculated gender data and its interpretation.

We have used an updated version of Teseo data downloaded during the last months of 2023 that includes 298,584 records on doctoral dissertations from 1977 to 2022. Records were extracted from the Website using scrapping techniques and downloaded to a local database. Entities such as individuals, institutions or subject matters were extracted using simple automatic procedures and then manually checked and normalized whenever possible. Individuals comprised in the dissertation records were linked to the dissertations in which they participated including information on their corresponding role (candidate, supervisor, co-supervisor, member of the examination committee, chair of the committee). Our data thus includes all the individuals that were involved in creation, supervision or examination of dissertations presented up until the end of 2022 and included in the public database up until November 2023.

We reproduced the original simple matching method based on INE data in the updated version of the database to assign gender to individuals within the entire Spanish doctoral system. Some refinements were used, including a name parser and the inclusion of a modest set of heuristics, aiming to identify gender from unstructured labels containing name and surname information. We also used a newer method based on the Cultural Consensus Theory Classifier described above (NQG). This method was well proven in other contexts, and our own tests with researchers based in Spain seem to indicate that its performance should be very good. Also, it seemed like a good candidate to reduce the uncertainty associated with unclassified data.

**Table 6** shows a comparison of gender distributions among the two methods. Considering only individuals for whom gender assignment has been achieved, the original percentages of female academic staff (37.3%) and the percentages obtained using the method based on cultural consensus theory (39.4%) are quite close. However, the category of unknow gender varies greatly in size between methods. Using only INE data leads to a 18.2% unknown gender rate, while the new method is unable to assign a gender to only a 2.5%. The difference is greater than expected, so we have closely examined this particular aspect in order to try understanding its possible effects.

| Gender (NQG) | Count | % total | % assigned gender | Gender (INE) | Count | % total | % assigned gender |
|---|---|---|---|---|---|---|---|
| Male | 274,035 | 59.1% | 60.6% | Male | 237,893 | 51.3% | 62.7% |
| Female | 178,541 | 38.5% | 39.4% | Female | 141,813 | 30.6% | 37.3% |
| Unknown | 11,449 | 2.5% | | Unknown | 84,319 | 18.2% | |

**Table 6**: Assigned gender counts and percentages for academics in the Teseo database, as determined by NQG and Sánchez-Jiménez 2023 methods.

The *unknown* category includes labels that could not be processed (because the proper name was not identified in the original label provided by Teseo), names that have no match in any of the data sources, and names for which the available information did not allow determining the gender with certainty. Among the 90,126 name tags that could not be classified by gender using the simple matching procedure 71,361 did not find a compatible name in the INE dataset. Following an exploratory review of a subset of 1100 tags, we concluded that many of these tags (around 62%) were not recognized because they used compound names, many of which included initials instead of the full name. Also, a relevant percentage (around 30%) corresponded to authors with foreign names. INE data provides information on names of residents (regardless of their origin), although names with low frequency are more likely to be missing. Finally, a small number of names (7.5%) included only initials, reversed the order of names and surnames, contained typing errors, or were extremely uncommon names. This reinforces our idea that the preprocessing step is crucial in obtaining good results. On the other hand, it also highlights the need to use sources with a less local focus for gender studies in science, even when the analyzed context is only national and the need to provide nontrivial solutions to the problem of compound names.

**Table 7** provides a more useful insight into determining the effect that the new method has had on gender data. The original *non-binary* category included academics whose names did not allow assigning a gender with certainty. Some names are deliberately ambiguous regarding gender, and some others are conditionally gendered according to cultural or even temporal settings. Deciding which of these reasons could be behind our inability to assign a gender would require a deeper understanding of how and why names are associated to genders. This is out of the scope of our work, and probably very difficult to tackle using today's most common approach to gender detection, which is based on given names. In our gender classification procedure, we have opted for a generic "unknown" category that includes both individuals with probably non-binary names (or weakly gendered names) and individuals for which sufficient data is not available, for whatever the reason. Only a 4.5% of the original "non-binary" names were assigned to the new NQG unknown category, while 40.7% of individuals were assigned the female gender and 54.8% were assigned the male gender. This proportion is similar to the overall population proportion detected by the two methods, so it does not seem concerning.

On the other hand, the original *unknown* category shows a more balanced distribution (41.1% female and 45.2% male) among the categories assigned with NQG. However, this does not align as well with the expected gender share. In other words, there seems to be a bias that makes women overrepresented in this category, which could potentially influence the overall data and thus modify the representation of gender-related phenomena studied through the original data. We believe that measuring and attempting to interpret the effects this might have is interesting, as many studies significantly rely on automated methods to detect or assign gender.

| Gender | Female (NQG) | % Fem. (NQG) | Male (NQG) | %Male (NQG) | Unknown (NQG) | % Unk. (NQG) | Total count | %Total |
|---|---|---|---|---|---|---|---|---|
| Female (INE) | 141,565 | 99.8% | 161 | 0.1% | 87 | 0.1% | 141,813 | 30.6% |
| Male (INE) | 2,348 | 1.0% | 235,523 | 99.0% | 22 | 0.0% | 237,893 | 51.3% |
| Non binary (INE) | 1,023 | 40.7% | 1,377 | 54.8% | 113 | 4.5% | 2,513 | 0.5% |
| Unknown (INE) | 33,605 | 41.1% | 36,974 | 45.2% | 11,227 | 13.7% | 81,806 | 17.6% |
| Total | 178,541 | 38.5% | 274,035 | 59.1% | 11,422 | 2.5% | 464,025 | 100.0% |

**Table 7**: Distribution of genders assigned by NQG over genders assigned using INE data

As data on the gender assigned to names varies over time, we first plotted the evolution of the gender balance figures during the entire period (1977-2022). **Figure 2** shows that the trends in the percentages of female academics over the population with an assigned gender have grown steadily in both cases. Counts using the original procedure based on INE data have visibly lower female percentages over the whole period, although the trends are very similar, and both series mimic each other for most of the time, showing a very high Pearson correlation (0.999, p-value < 0.001). This would imply that names that belonged to the unknown category change to some extent our understanding of the phenomena that we are studying. That is, gender imbalance seems in general slightly less accentuated than we thought, although the trends are basically identical.

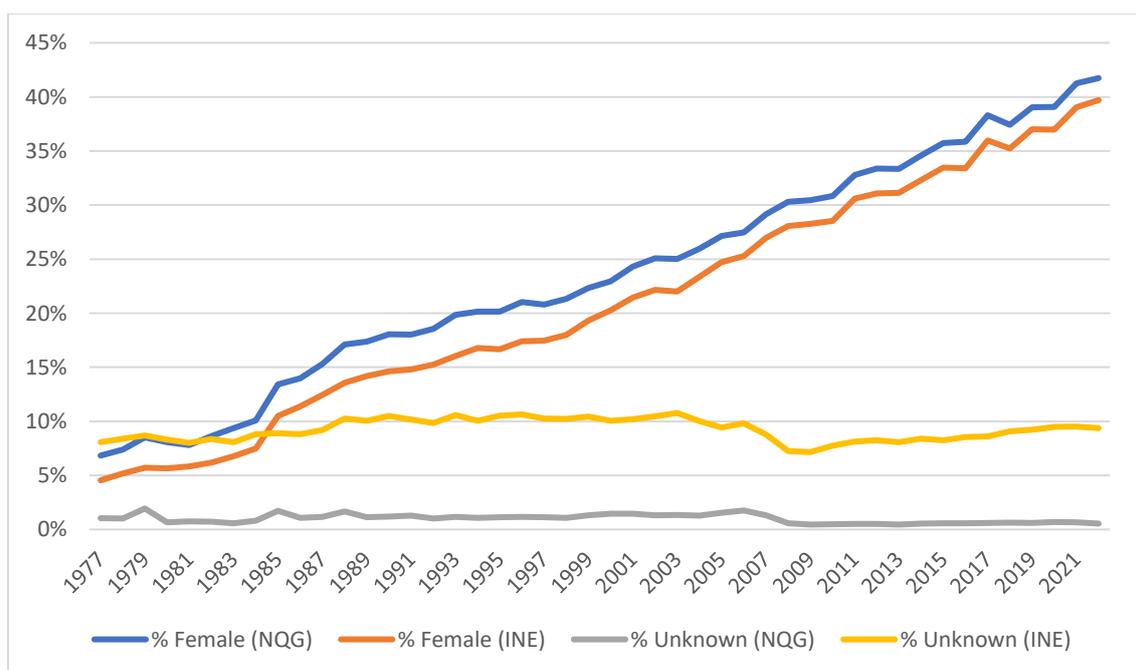

**Figure 2**: Evolution of % female academics in the Teseo database as detected by two different methods.

However, this data does not address the question of why the original method underrepresented women's participation. Considering the high percentage of compound names detected in the sample of 1,100 labels from the unknown category, we systematically analyzed the data for the whole collection. Approximately 150,000 individuals in the dataset have compound names. Of these, around 55,000 do not have an exact match in the INE data. According to the NQG method's classification, 2.67% would remain without an assigned gender, while 52.85% would be classified as female and 44.48% as male. Additionally, many compound names used initials to abbreviate the names (e.g., M. Dolores instead of María Dolores, F. José instead of Francisco José), which significantly affected women's names (11,432) more than men's names (4,243). This rendered name labels unrecognizable for the old method, but not for the new one. This issue has been gradually corrected over time, and although the difference persists, it is proportionally much smaller in the last years. This phenomenon would explain at least a substantial part of the divergence in the percentages of females in the two data series.

In **Table 8**, we can see a summary of the first-level UNESCO categories, which constitute the most substantial thematic divisions in the database. This classification system is very outdated and to a large extent, incompatible with classification systems used in modern evaluation processes or scientometric endeavors. This is reflected in the prevalence of subject matters from the Social Sciences and Humanities, as well as the lack of definition of the subject matters in the Biomedical branch of science, aside from the absence of relevant subject areas like Nursing, or other more specific scientific categories. Even though, this can be used as an approximation to the distribution of scientific interest in Spanish universities.

| Subject | Male (INE) | Female (INE) | Male (NQG) | Female (NQG) | % Fem. (INE) | % Fem. (NQG) | N. of theses | Diff. |
|---|---|---|---|---|---|---|---|---|
| **Medical Sciences** | 139,744 | 28,492 | 146,661 | 36,550 | 16.9% | 19.9% | 27,116 | 3.0% |
| **Chemistry** | 111,498 | 41,857 | 117,457 | 50,509 | 27.3% | 30.1% | 24,528 | 2.8% |
| **History** | 82,852 | 27,490 | 87,336 | 33,658 | 24.9% | 27.8% | 18,680 | 2.9% |
| **Logic** | 83,889 | 16,975 | 88,698 | 21,627 | 16.8% | 19.6% | 16,076 | 2.8% |
| **Life Sciences** | 75,189 | 22,174 | 79,174 | 27,650 | 22.8% | 25.9% | 15,702 | 3.1% |
| **Physics** | 68,461 | 13,127 | 73,131 | 16,441 | 16.1% | 18.4% | 13,200 | 2.3% |
| **Ethics** | 54,719 | 22,436 | 57,334 | 26,024 | 29.1% | 31.2% | 12,489 | 2.1% |
| **Technological Sci.** | 64,314 | 9,964 | 68,077 | 13,082 | 13.4% | 16.1% | 11,719 | 2.7% |
| **Psychology** | 41,557 | 21,884 | 43,691 | 25,888 | 34.5% | 37.2% | 10,460 | 2.7% |
| **Mathematics** | 37,586 | 5,536 | 39,800 | 7,248 | 12.8% | 15.4% | 6,901 | 2.6% |
| **Economic Sciences** | 36,527 | 6,581 | 38,226 | 8,511 | 15.3% | 18.2% | 6,863 | 2.9% |
| **Science of Arts and Letters** | 30,957 | 10,278 | 32,748 | 12,680 | 24.9% | 27.9% | 6,811 | 3.0% |
| **Pedagogy** | 25,443 | 12,552 | 26,789 | 15,076 | 33.0% | 36.0% | 6,342 | 3.0% |
| **Juridical Sci. & Law** | 31,954 | 6,151 | 34,026 | 7,598 | 16.1% | 18.3% | 6,331 | 2.1% |
| **Linguistics** | 24,013 | 12,223 | 25,399 | 15,017 | 33.7% | 37.2% | 6,120 | 3.4% |
| **Sociology** | 24,944 | 10,631 | 26,218 | 12,576 | 29.9% | 32.4% | 6,021 | 2.5% |
| **Philosophy** | 24,909 | 6,046 | 26,176 | 7,124 | 19.5% | 21.4% | 5,201 | 1.9% |
| **Agricultural Sci.** | 23,133 | 6,098 | 24,678 | 7,500 | 20.9% | 23.3% | 4,627 | 2.4% |
| **Earth and Space Sciences** | 21,573 | 4,190 | 22,816 | 5,520 | 16.3% | 19.5% | 4,124 | 3.2% |
| **Anthropology** | 14,836 | 5,460 | 15,573 | 6,637 | 26.9% | 29.9% | 3,370 | 3.0% |
| **Geography** | 12,942 | 3,922 | 13,622 | 4,944 | 23.3% | 26.6% | 2,852 | 3.4% |
| **Political Science** | 12,099 | 3,031 | 12,741 | 3,777 | 20.0% | 22.9% | 2,517 | 2.8% |
| **Astronomy & Astrophysics** | 4,535 | 951 | 4,928 | 1,134 | 17.3% | 18.7% | 893 | 1.4% |

| Demographics | 2,332 | 777 | 2,464 | 940 | 25.0% | 27.6% | 520 | 2.6% |

**Table 8**: Distribution of genders assigned by NQG and genders assigned using INE data over UNESCO subject matters. Percentages are relative to the total number of individuals with an identified gender.

Differences in the percentages of females detected by the two methods are visible but discreet. The macro-averaged difference for the 24 first-level subjects is 2.7%, almost the same as for the 227 second-level subjects (2.6%) and slightly higher than the 2.3% for the 2141 third-level subjects, which refer to much more specific topics. Since the differences sometimes invert (there are more female academics according to the method that uses INE data than to NQG), a more useful metric would be the coefficient of variation. The first and second-level subjects accumulate variations between the two methods that are still discreet (0.17 and 0.28, respectively), but the cumulative level of variations for the third-level subjects is much higher (0.66). This might be indicative of a resolution problem for smaller units of analysis, although results at an aggregated level probably won't be affected significantly.

In summary, the results provided by the two methods allow us to reach compatible conclusions at high or medium levels of aggregation, although divergences may appear in more specific contexts. The method based on cultural consensus offers better coverage for foreign individuals, who have significant but more limited representation in the data on which the original method is based. Additionally, its ability to work with low frequency compound names and incomplete data helps to significantly reduce the percentage of individuals without assigned gender and, secondarily, to avoid biases and artifacts in data.

**Assessing gender imbalance in doctoral education**

Once we have described the effects of using a more comprehensive and refined procedure on our ability to discern the gender of individuals, we can attempt to reassess the existence of gender imbalance in the doctoral context in Spain. Taking into account the slightly fewer than two million references of individuals participating in the doctoral process in the Teseo records (roughly 6.7 individuals per thesis, including the author, supervisor/s and members of the committee) we obtain a distribution that indicates a very clear gender imbalance. 70.14% of the references correspond to men, compared to 28.97% to women, and 0.88% without an assigned gender. If we compare this proportion to the roughly 60/40% balance of male/female individuals in the database (**table 6**) we can conclude that not only have been fewer female actors involved, but they have also had a proportionally lower presence in doctoral processes. The situation has improved overtime and **figure 2** provides a more relevant reference for understanding the current context and trends. For the most recent year, women's participation in the doctoral process had improved, but was significantly lower than that of men, at 41.7% compared to 58.3% male participation.

**Table 8** does also provide a global perspective on the distribution of female and male participation over disciplines. In the field of Arts and Humanities (History, Logic, Ethics, Science of Arts and Letters, Linguistics, and Philosophy), many disciplines show female participation percentages significantly higher than the average of 24.9%. The most notable case is Linguistics, although only Logic and Philosophy are below the average (regardless of the gender classification method). In the STEM areas (Chemistry, Life Sciences, Physics, Technological Sciences, Mathematics, Agricultural Sciences, Earth and Space Sciences, Astronomy & Astrophysics) and Medicine, the opposite occurs. Except for Chemistry, which is visibly above the average (30.1%), and Life Sciences (25.9%), all other disciplines have much lower female participation. The severity of the problem within the STEM context has been described by many authors in both educational and research contexts (White, **2004;** Blickenstaff**, 2005;** Blackburn, **2017**). Data clearly shows that the problem is also concerning at the intersection of education and research. Lastly, the social sciences (Psychology, Economic Sciences, Pedagogy, Juridical Sciences & Law, Sociology, Anthropology, Geography, Political Science, and Demographics) are closer to the average than the other two contexts (28.7%), although with disciplines at both extremes. Psychology and Pedagogy both exceed 36% female participation, while Economic Sciences and Juridical Sciences only slightly surpass 18%

This distribution does not take into account the evolution of the system over time, though. **Figure 3** provides a general overview of how the main disciplines have evolved over time and shows that the structural distribution of female participation by subject has changed over the years. Overall, the four main branches of science (Medical Sciences, STEM, Social Sciences, and Humanities) have significantly evolved over the years, reducing the gender imbalance. Medical Sciences likely exhibit the most distinctive evolution, as it had the lowest percentage of women during the 1980s, 1990s, and much of the 2000s, but is currently the second branch, just behind Social Sciences (43.6% vs. 46.3% in 2022). Overall, the trend leans decisively towards gender balance, though important nuances remain. In STEM, Chemistry and Life Sciences not only have higher percentages of females but also tend to diverge from other disciplines that show a much slow increase in female participation. In Humanities, fields like Linguistics or even Arts & Letters have seen increasing female percentages, even though they have already surpassed 50%, while in History or Ethics, parity may not be reached in the short term. In Social Sciences, on the other hand, although some smaller fields show worse trends, all major scientific fields (including Economic Sciences and Juridical Sciences) seem to be progressing towards parity in relatively few years if current trends continue, or have already reached that point.

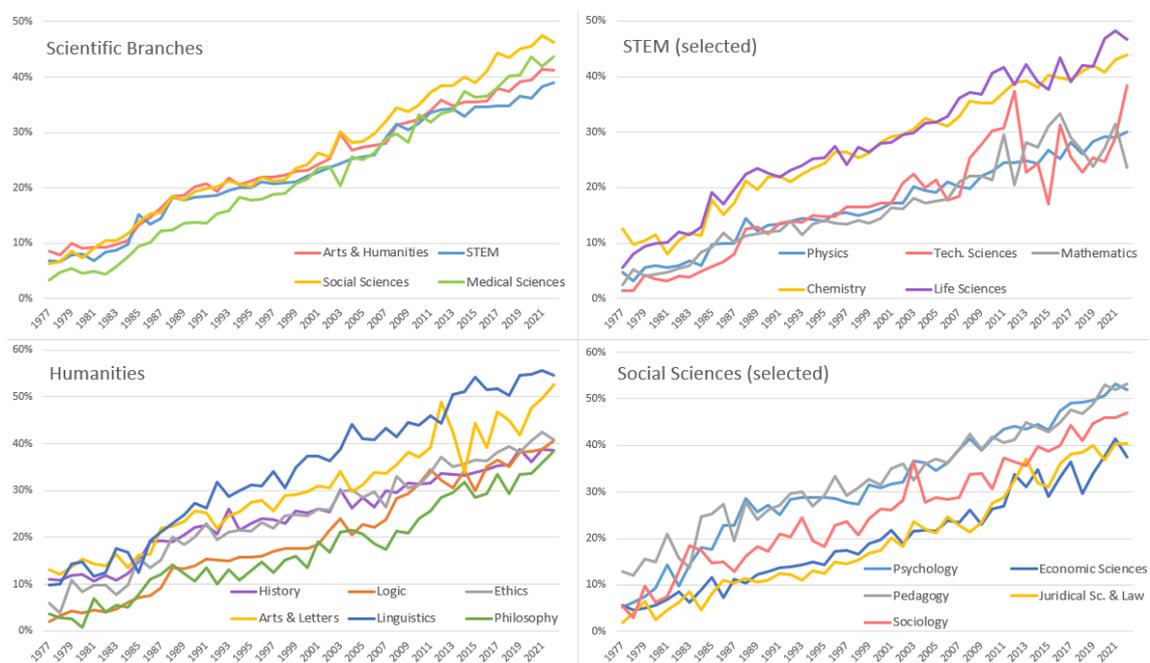

**Figure 3**: Percentage of females participating in dissertation examinations in any role. Evolution by scientific area (selected) and branch. Genders assigned using NQG method.

None of the previous analyses distinguish between the usual roles of candidate, committee member, supervisor, and committee chair, but important differences can be observed when considering these roles in conjunction with the gender of the participants. **Figure 4** shows a distribution of female participation by roles that is both interesting and worrying. For several years, female candidates have been slightly above or below 50% of individuals who successfully defend their doctoral theses. However, the proportions of women in more senior roles are progressively lower. Committee members have not yet reached parity (46.1%) for the most recent year available (2022), although the trend clearly indicates that they will soon. Supervisors are much further away (36%), and the trend towards parity appears somewhat slower. Finally, the most senior role, that of committee chair, has the lowest female participation, at 33.3%.

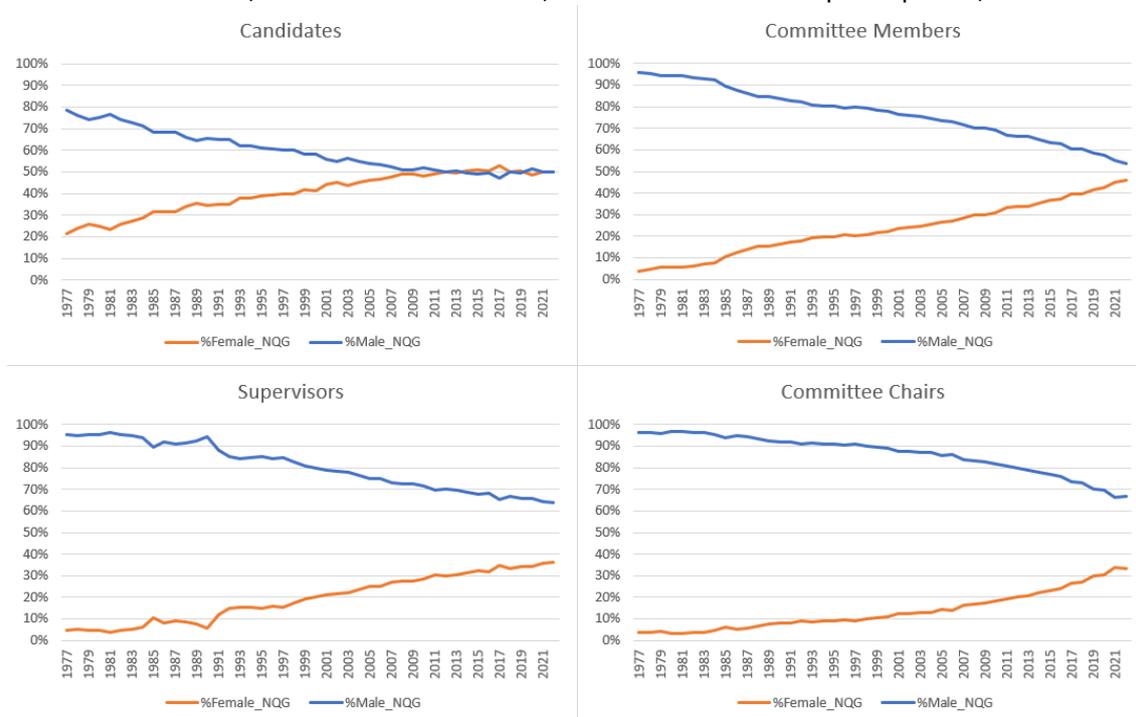

**Figure 4**: Evolution of % female academics in the Teseo database according to their role. Genders assigned using NQG method.

In general terms we can say that there is a clear gender inequality in the participation of the individuals involved in the doctoral system. This gap shows an overall trend towards closing, although the current situation and pace of evolution vary significantly depending on the role of the individuals and the scientific area. Specifically, it appears evident that there are many scientific fields in which female participation already exceeds 50%, and many others where this level will be reached in the coming years. However, we can confirm the existence of two concerning aspects that the original study found. Firstly, the existence of disciplines where the gap is still significant and will take time to close at the current rate, and secondly, the significant differences between roles based on seniority. This can be associated to the lack of parity in leadership roles, be it because the existence of a leaky pipeline, a glass ceiling, or a combination of both, as suggested by Surawicz (2016).

## Discussion and Conclusions

Gender has become an increasingly important dimension in the study of scientific activity, and many researchers have devoted time and effort to the task of analyzing, detecting, describing and characterizing gender imbalances in science. These efforts heavily rely on the quality of the methods employed, and thus reflecting over the accuracy of the gender assigning process is relevant to the whole set of studies that deal with gender studies in science.

We have attempted to create an evaluation procedure for gender assignment methods that is reasonably straightforward and based on well-known metrics and data freely provided by other researchers. We believe that the results of the evaluation process can offer valuable insights into the performance of various well-known options for gender assignment. At the same time, it becomes evident that certain starting conditions of the multiple existing methods should be critically analyzed when determining the best available option. Temporal and geographical coverage, as well as the overall breadth of available sources, appear to be crucial for achieving good results, even in local contexts such as the Spanish science and technology system.

Transparency and the free availability of data also seem to be important criteria, facilitating the repeatability of research results and, indirectly, the progress of studies on the issue. Lastly, since the performance levels shown by widely used systems are very similar, the availability of free options with good access conditions seems to provide a clear path for addressing the issue of gender assignment to individuals in the context of scientific activity and of course in other contexts. We believe that the method based on Cultural Consensus Theory that we selected can be used successfully for similar purposes on a generalized context given its performance and the wide array of data sources it's based on.

We have employed this new method to classify scholars in Teseo database, for which a recent study on gender imbalance around doctoral theses had been published. After recalculating the figures that allowed the original analyses, we can observe how the use of a more sophisticated method with a broader dataset and better preprocessing significantly reduced the number of academics for whom gender was unknown. This allows us to expand the coverage of the study, but it also raises the question of whether bias was occurring within the data with unknown gender. It indeed seems that an artifact of data that was not detected using the older method was creating a picture in which female participants were even more underrepresented.

The impact of this on interpreting the research results is modest if we look at the general distributions and trends in data. Gender imbalance is somewhat smaller than we thought, but it is still very much there. We should be cautions though when using smaller aggregation levels. In these cases, choosing a method with higher performance may be important for gender studies. Sectorial studies or those in which the literature selection process is constrained by special circumstances (narrow production windows, specific methodological approaches, ad-hoc literature searches) should probably use high-quality approaches or manual labeling to reduce the chances of an unrefined method introducing hidden biases in the data labeling process.

When we apply the new method to the Teseo data, we can confirm that there is a clear gender imbalance in the system. Nevertheless, it is also evident that the general participation of women has been progressively increasing in all analyzed facets. Moreover, if current trends continue, it is expected that the system as a whole will clearly move away from this imbalance. On the other hand, specific problems associated with STEM fields, as well as some fields in the Arts & Humanities and Social Sciences should still be a source of concern. Additionally, the lack of balance in seniority roles appears to be a marked pattern that still deserves the attention of researchers and policymakers.

---

[i] Gender labels for Spanish affiliated researchers. Available online: https://zenodo.org/doi/10.5281/zenodo.11243211